\documentclass[twocolumn,superscriptaddress,longbibliography]{revtex4-1}
\usepackage{amsmath}
\usepackage{amssymb}
\usepackage{bm}
\usepackage{braket}
\usepackage{color}
\usepackage{varwidth}
\usepackage{dcolumn}
\usepackage[breaklinks,colorlinks=true,linkcolor=blue,urlcolor=cyan,citecolor=blue]{hyperref}
\usepackage{graphicx}
\usepackage{algorithm}
\usepackage{algorithmicx}
\usepackage{algpseudocode}
\usepackage{empheq}
\algdef{SE}[DOWHILE]{Do}{doWhile}{\algorithmicdo}[1]{\algorithmicwhile\ #1}
\newtheorem{thm}{Theorem}

\begin{document}
\title{Reducing rejection exponentially improves Markov chain Monte Carlo sampling}

\author{Hidemaro Suwa}
\email{suwamaro@phys.s.u-tokyo.ac.jp}
\affiliation{Department of Physics, The University of Tokyo, Tokyo 113-0033, Japan}

\begin{abstract}
  The choice of transition kernel critically influences the performance of the Markov chain Monte Carlo method.
  Despite the importance of kernel choice, guiding principles for optimal kernels have not been established. 
  Here, we propose a one-parameter rejection control transition kernel that can be applied to various Monte Carlo samplings and demonstrate that the rejection process plays a major role in determining the sampling efficiency.
  Varying the rejection probability, we examine the autocorrelation time of the order parameter in the two- and three-dimensional ferromagnetic Potts models.
  Our results reveal that reducing the rejection rate leads to an exponential decrease in autocorrelation time in sequential spin updates and an algebraic reduction in random spin updates.
  The autocorrelation times of conventional algorithms almost fall on a single curve as a function of the rejection rate.
  The present transition kernel with an optimal parameter provides one of the most efficient samplers for general cases of discrete variables. 
\end{abstract}

\date{\today}
\maketitle

\section{Introduction}
\label{sec:intro}
The choice of the transition kernel, or the transition probability, is paramount to enhancing the efficiency of the Markov chain Monte Carlo (MCMC) method~\cite{Newman1999,Landau2005}. 
An optimally tailored transition kernel can expedite convergence to the target distribution and markedly decrease autocorrelation, thereby enhancing the overall effectiveness of the sampling process.
The transition kernel is chosen to ensure global balance, which guarantees the invariance of a target distribution from which the states are sampled. 
Global balance is not a tight condition, leaving much room for optimization. 
In many applications, detailed balance, or reversibility, is additionally imposed as a sufficient condition for global balance.
Although detailed balance is not necessary for sampling from a stationary distribution~\cite{Robert2004}, it becomes easy to set an appropriate transition kernel thanks to the reversibility.
Many reversible samplers have been proposed and tested, such as the Metropolis algorithm~\cite{Metropolis1953}, the heat bath algorithm~\cite{Creutz1980,Geman1984}, the Metropolized Gibbs sampler~\cite{Liu1996}, and the iterative Metropolized Gibbs sampler~\cite{Frigessi1992}. 
Recently, several nonreversible samplers have been developed to achieve faster convergence and smaller statistical errors, such as the Suwa--Todo algorithm~\cite{SuwaT2010}, lifting techniques~\cite{TuritsynCV2011,FernandesW2011,Schram2015,Vucelja2016,Faizi2020}, and the event-chain Monte Carlo method~\cite{BernardKW2009,MichelKK2014}.
However, guiding principles for optimal transition kernels have not been established.
For a finite state space spanned by discrete variables, the autocorrelation time can be represented by the eigenvalues of the transition matrix~\cite{Frigessi1992,Peskun1973}.
The optimization problem of the transition matrix is equivalent to controlling eigenvalues by tuning matrix elements, which is highly nontrivial for a huge matrix typical of statistical mechanical problems.

Rejection minimization algorithms have been shown to outperform Metropolis and heat bath algorithms in several models~\cite{SuwaT2010,SuwaT2012,Suwa2021}.
Because it should be favorable to allow exploring the state space quickly in the MCMC update, the rejection process, leaving the configuration unchanged, naturally deteriorates the sampling efficiency. 
It is mathematically proved that a reversible transition matrix with larger off-diagonal elements produces a smaller asymptotic variance, that is, a higher sampling efficiency~\cite{Peskun1973}.
This mathematical theorem implies that decreasing diagonal elements, namely rejection probabilities, provides a better sampler.
However, this theorem does not apply to nonreversible cases.
Detailed studies on the relationship between the rejection rate and the sampling efficiency are lacking.
Thus, it is desirable to develop an approach to controlling the rejection probability and conduct a systematic study of the effect of the rejection process.

Here, we propose a one-parameter transition kernel that controls the rejection probability in the MCMC method. 
The single parameter determines the shift in the cumulative distribution of the configuration weights. 
The present algorithm always holds the global balance and additionally ensures the detailed balance with a particular parameter choice.
We can easily find optimal parameters to minimize the rejection probability. 
Reviewing the optimization problem and previous approaches in Sec.~\ref{sec:prob}, we present our approach in a graphical picture and an analytical expression in Sec.~\ref{sec:one}.
The effect of rejection on sampling efficiency is investigated for the square lattice and the simple cubic lattice ferromagnetic Potts models in Sec.~\ref{sec:result}.
As the rejection probability is reduced, the sampling efficiency improves exponentially in the sequential spin update and algebraically in the random spin update.
The autocorrelation times of the reference algorithms almost collapse onto a single curve as a function of the rejection rate, indicating that the rejection process is a major factor in determining the sampling efficiency of the MCMC method.
The present paper is concluded in Sec.~\ref{sec:con}.

\section{Probability optimization}
\label{sec:prob}
We here briefly review the optimization problem of the transition kernel and previous approaches.
Let us consider a finite state space $S$ spanned by discrete state variables and a Monte Carlo sampling from a target distribution $\pi: S \to \mathbb{R}$.
Our task is to optimize the transition kernel, or the transition matrix $P$, under the condition of distribution invariance, namely the global balance~\cite{Robert2004}:
\begin{equation}
    \pi_i = \sum_{j\in S} \pi_j P_{ji} \qquad \forall i \in S,
    \label{eq:gb}
\end{equation}
where $\pi_i$ is the weight, or the measure, of state $i \in S$, $P_{ji}$ is the transition probability from $j$ to $i$, and $\sum_i P_{ji} = 1$.
If a transition matrix satisfies the detailed balance condition,
\begin{equation}
\pi_j P_{ji} = \pi_i P_{ij} \qquad \forall i,j \in S,
\label{eq:dbc}
\end{equation}
it is called reversible with respect to $\pi$.
Detailed balance is sufficient for global balance but not necessary.

In the MCMC method, one needs to reduce the autocorrelation between samples.
The autocorrelation function~\cite{Newman1999,Landau2005} of an observable $\mathcal{O}$ is defined by
\begin{equation}
    A_{\mathcal{O}}(t)=\frac{\langle \mathcal{O}_{i+t} \mathcal{O}_i \rangle - \langle \mathcal{O} \rangle^2}{\langle \mathcal{O}^2 \rangle - \langle \mathcal{O} \rangle^2 },
    \label{eq:A}
\end{equation}
where $\mathcal{O}_t$ is the sample of the observable at the $t$th Monte Carlo step, and $\langle \cdot \rangle$ denotes the Monte Carlo average.
This function \eqref{eq:A} becomes almost independent of $i$ after thermalization or burn-in.
In many cases, the autocorrelation function decays exponentially for large $t$:
$A_{\mathcal{O}}(t) \sim e^{-t/\tau_{{\rm exp},\mathcal{O}}}$, where $\tau_{{\rm exp},\mathcal{O}} = - 1/\ln |\lambda_2|$ is the exponential autocorrelation time, given by the second largest eigenvalue $\lambda_2$ of the transition matrix under the assumption that the observable $\mathcal{O}$ has a nonzero projection onto the corresponding eigenstate.
The sampling efficiency of the MCMC method can be quantified by the integrated autocorrelation time:
\begin{equation}
    \tau_{{\rm int},\mathcal{O}} = \frac{1}{2} + \sum_{t=1}^\infty A_{\mathcal{O}}(t).
    \label{eq:tauint}
\end{equation}
For reversible cases, it is straightforward to obtain the relation between the integrated autocorrelation time and the eigenvalues of the transition matrix~\cite{Frigessi1992,Peskun1973}.
Thus, the transition kernel optimization problem is equivalent to controlling the eigenvalues of a transition matrix by tuning the matrix elements~\cite{Chen2012,Suwa2014}, which is highly nontrivial for a large matrix.

Fortunately, Peskun's theorem~\cite{Peskun1973} provides a practical route to improving reversible Markov chains.
Let us define an order of matrices: $P_2 \geq
P_1$ for any two transition matrices if each of the off-diagonal
elements of $P_2$ is greater than or equal to the corresponding
off-diagonal elements of $P_1$. 
The following is Theorem
2.1.1 of Ref.~\cite{Peskun1973}.

\begin{thm}[Peskun]
Suppose each of the irreducible transition matrices $P_1$ and $P_2$ is
reversible for a same invariant probability distribution $\pi$. If
$P_2 \geq P_1$, then, for any $f$,
\begin{equation}
v(f, P_1, \pi ) \geq v( f, P_2, \pi ),
\end{equation}
where
\begin{equation}
v(f,P,\pi) = \lim_{M \rightarrow \infty} M\, {\rm Var}(\hat{I}_M),
\end{equation}
and $\hat{I}_M = \sum_{i=1}^M f(x_i)/M$ is an estimator of $I =
E_{\pi}[f]$ using $M$ samples, $x_1, x_2, ..., x_M$, of the Markov
chain generated by $P$.
\label{theo:p}
\end{thm}
\noindent
Thus, larger off-diagonal elements yield a smaller asymptotic variance $v$, leading to a smaller statistical error.

According to this theorem, a modified Gibbs sampler was proposed called the
Metropolized Gibbs sampler~\cite{Liu1996,Frigessi1992}. 
In the standard Gibbs sampler~\cite{Creutz1980,Geman1984}, or the heat bath algorithm, the next state is chosen with a probability proportional to its weight independently of the current state.
In the Metropolized Gibbs sampler, a state is proposed with a probability proportional to its weight, excluding the current state~\cite{Loison2004}.
The proposed state is then accepted or rejected by the Metropolis algorithm (filter).
The resulting transition probability, which is the product of the proposal and the acceptance probabilities, is given by
\begin{equation}
P_{ji}^{\rm MG} = \min \left( \frac{\pi_i}{ 1 - \pi_j}, \frac{\pi_i}{ 1 - \pi_i} \right) \qquad \forall i \neq j,
\label{eq:mg}
\end{equation}
where the states are ordered such that $\pi_1 \leq \pi_2 \leq \dots \leq \pi_n$, and $n$ is the number of possible states.
Here, we assume that the weights are normalized: $F_n = \sum_{i=1}^n \pi_i = 1$.
Otherwise, the denominators $1- \pi_j$ and $1 - \pi_i$ in Eq.~\eqref{eq:mg} are replaced with $F_n - \pi_j$ and $F_n - \pi_i$, respectively.
Normalizing the weights is nontrivial for a large $n$ but easy for a small $n$, such as in updating a local variable.
For example, as studied in the following, we set $n=q$ for the single spin update of the $q$-state Potts model.
The diagonal elements are determined from Eq.~\eqref{eq:mg} and the probability conservation $\sum_i P_{ji}=1$.
It then follows that $P^{\rm MG} \geq P^{\rm G}$, where $P^{\rm G}$ is the transition matrix of the Gibbs sampler.
Therefore, the Metropolized Gibbs sampler is more efficient than the Gibbs sampler.
Note that the Metropolized Gibbs sampler reduces to the Metropolis algorithm for $n=2$.

Let us define the rejection process as the process in which the next state is the same as the current state.
The rejection probability is thus given by the diagonal elements of the transition matrix.
The above transition probabilities~\eqref{eq:mg} always lead to $P_{11}^{\rm MG}=0$; the rejection probability for the smallest-weight state becomes zero, which is called the Metropolization for this state.

Interestingly, we can iterate this Metropolization scheme for other states and set $P_{ii}=0$ for $1\leq i \leq n-1$~\cite{Frigessi1992,Pollet2004}.
The resultant transition probabilities are given by
\begin{equation}
P_{ji}^{\rm IMG} = 
\begin{cases}
    y_i & (i < j)\\
    \frac{\pi_i}{\pi_j}y_j & (i > j) \\ 
    1 - \sum_{k=1}^{n-1} y_k & (i=j=n) \\
    0 & (i=j<n),
\end{cases}
\label{eq:PIMG}
\end{equation}
where $y_1 = \pi_1/(1-\pi_1)$ and
\begin{equation}
    y_j = \frac{1 - \sum_{k=1}^{j-1}y_k}{1 - \sum_{k=1}^j \pi_k} \pi_j \qquad (1 < j < n),
\end{equation}
which we call the iterative Metropolized Gibbs sampler.
Here we again assume that the weights are normalized.
This iterative sampler is generally more efficient than the simple Metropolized Gibbs sampler: $P^{IMG} \geq P^{MG}$.

Meanwhile, a rejection-minimization algorithm~\cite{SuwaT2010} was proposed using the geometric allocation approach, which we call the Suwa–Todo algorithm.
This algorithm and related rejection-minimization approaches were tested for several physical models and shown to be more efficient than the Metropolis algorithm, the heat bath algorithm, and even the iterative Metropolized Gibbs sampler~\cite{SuwaT2010,SuwaT2012}.
The Suwa--Todo algorithm produces a nonreversible Markov chain, so Peskun's theorem does not apply to it.

Questions here are how much the rejection probability influences the efficiency of the MCMC method and whether the rejection process plays a major role in determining the efficiency for both reversible and non-reversible samplers.
The primary purpose of the present paper is to answer these questions.

\section{One-parameter rejection control transition kernel}
\label{sec:one}
To study how the rejection process affects the efficiency of the MCMC method and to provide a flexible optimization approach, we propose a one-parameter transition kernel that enables control of the rejection probability.
Our approach is an extension of the geometric allocation approach to optimize stochastic flows between possible states~\cite{SuwaT2010,TodoS2013}.

\begin{figure}
  \begin{center}
\includegraphics[bb=0 0 1418 1292,width=0.7\columnwidth]{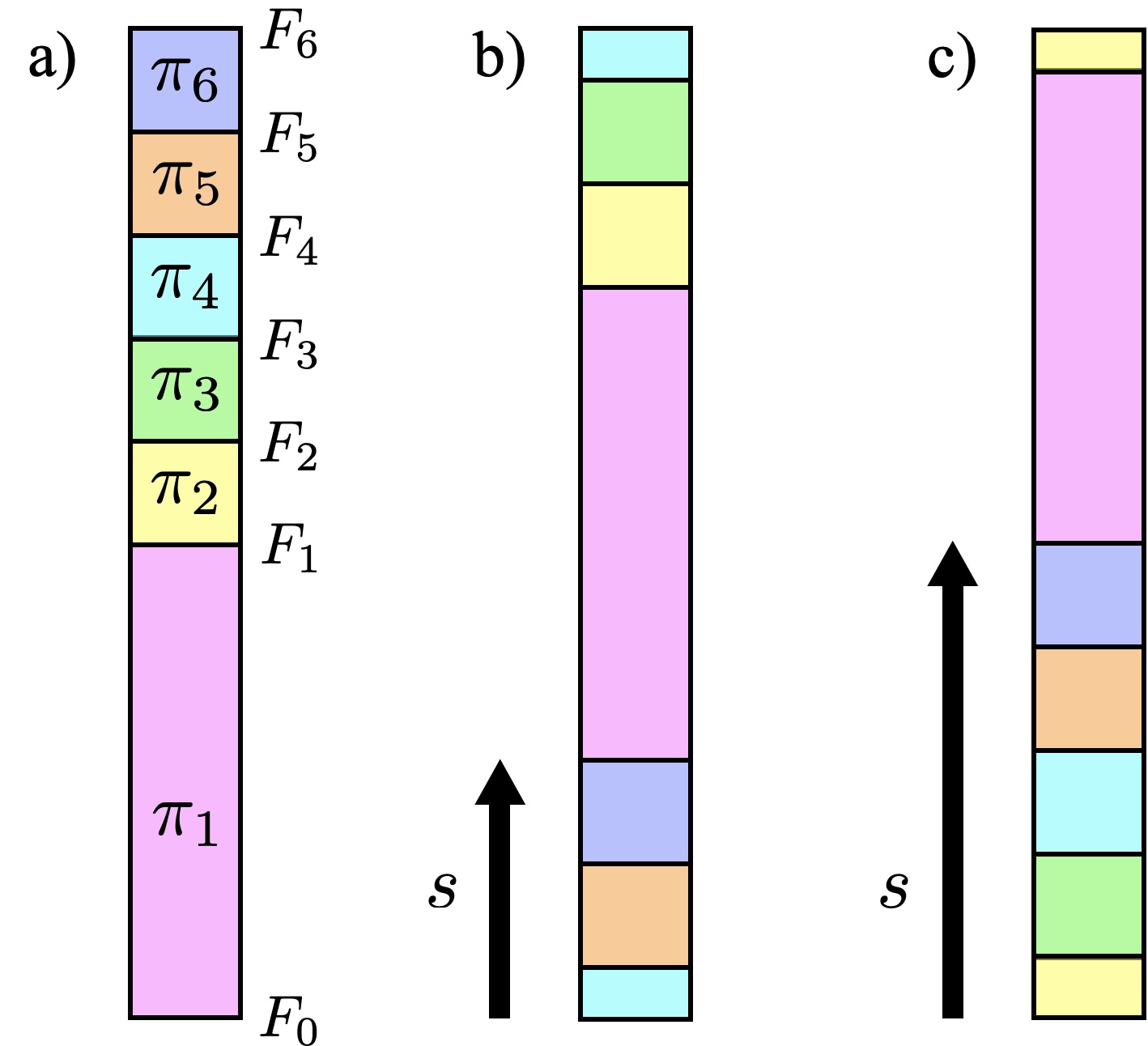}
\caption{\label{fig:shift}(a) Example of the weight tower for $n=6$ and (b) a periodic shift of the tower. 
$\pi_i$ is the weight of state $i$, and $F_i$ is the cumulative distribution.
The parameter $s$ represents the shift amount of the weight tower.
The stochastic flows and the transition probabilities are determined from the overlap between the original and shifted towers for each state. 
See the main text for the details.
(c) When $s=\max_i \pi_i$, this approach reduces to the Suwa–Todo algorithm~\cite{SuwaT2010}.
}
  \end{center}
\end{figure}

We first calculate the cumulative weight distribution:
\begin{equation}
 F_i= \sum_{j=1}^i \pi_j \qquad (1 \leq i \leq n)
\end{equation}
and $F_0=0$. 
Here, the order of the states is arbitrary. 
Calculating the cumulative distribution is nontrivial for a huge $n$ but easy for a small $n$, such as in the update of local variables.
Let us consider the sequence of $F_i$, or the weight tower, as illustrated in Fig.~\ref{fig:shift}~(a) for $n=6$.
We then consider shifting the tower by a certain amount, as in Fig.~\ref{fig:shift}~(b).
The shift is periodic; the weight shifted over the top of the tower is allocated to the lowest part of the tower.

Using this weight shift, we determine the transition probability from the overlap between the original and shifted towers.
It is convenient to define the stochastic flow from $i$ to $j$: $v_{ij}= \pi_i P_{ij}$.
Let $s$ denote the amount of the shift, $0 < s < F_n$.
We consider the overlap between the ranges $[F_{j-1}, F_j]$ and $[F_{i-1} + s, F_i + s]$. 
The stochastic flows $v_{ij}$ are set to the overlap for $i=1,2$ and $3$ in the case of Fig.~\ref{fig:shift}(b).
If $F_i + s > F_n$, we take into account an additional overlap between the ranges $[F_{j-1}+F_n, F_j+F_n]$ and $[F_{i-1} + s, F_i + s]$, assuming that the tower is periodic, for example, for $i=4,5$ and $6$ in Fig.~\ref{fig:shift}(b).
If a shifted weight range crosses the top of the original tower, nonzero flows are allocated from the overlaps in the above two cases, such as for $i=4$ in Fig.~\ref{fig:shift}(b) and $i=2$ in Fig.~\ref{fig:shift}(c).
We obtain the analytical expression of the stochastic flow determined by the periodic shift $s$:
\begin{eqnarray}
  v_{ij} &=& \max(0, \ \min( \Delta_{ij}, \ \pi_i + \pi_j - \Delta_{ij}, \ \pi_i, \ \pi_j)) \label{v_ij} \\
  && \hspace{-3.5mm} + \max(0, \ \min( \Delta_{ij} - F_n, \ \pi_i + \pi_j + F_n - \Delta_{ij}, \ \pi_i, \ \pi_j)) \notag,
\end{eqnarray}
where
\begin{equation}
\Delta_{ij} = F_i - F_{j-1} + s. \label{delta}
\end{equation}
The first term of Eq.~\eqref{v_ij} is the overlap between the ranges $[F_{j-1}, F_j]$ and $[F_{i-1} + s, F_i + s]$, and the second term is the overlap between the ranges $[F_{j-1}+F_n, F_j+F_n]$ and $[F_{i-1} + s, F_i + s]$.
See~\ref{App:v} for the derivation of the analytical expression.
This expression allows us to easily implement the present method.
The transition probability is then set to $P_{ij}=v_{ij}/\pi_i$.

Clearly, each range of $\pi_i$ in the original tower is covered by the shifted tower due to the periodic shift.
Our algorithm always satisfies the following condition:
\begin{equation}
    \pi_i = \sum_j v_{ji} \qquad \forall i \label{eq:gb2},
\end{equation}
equivalent to the global balance~\eqref{eq:gb}.
Let us consider using this algorithm for the update of local variables consisting of a many-body system, as in the single spin update of the Ising model.
Then assume that the number of local variables is $N$, and that the state is updated from $j$ to $i$ after updating the local variables one by one: $j \ (=k_0) \to k_1 \to k_2 \to \cdots \to k_{N-1} \to i\ (=k_N)$, where $k_\ell$ $(\ell=1,2, \dots, N-1)$ are intermediate states.
Let $P^{(\ell)}$ be a transition matrix in the update of the $\ell$-th local variable.
From Eq.~\eqref{eq:gb2} and $v_{ij}= \pi_i P_{ij}$, for any $i$, we can show
\begin{align}
  \pi_i &= \sum_{k_{N-1}} v_{k_{N-1}i} \notag\\
  &= \sum_{k_{N-1}} \pi_{k_{N-1}} P_{k_{N-1}i}^{(N)} \notag\\
  &= \sum_{k_{N-2}k_{N-1}}  \pi_{k_{N-2}} P_{k_{N-2}k_{N-1}}^{(N-1)} P_{k_{N-1}i}^{(N)} \\
  &= \sum_j \sum_{k_1 k_2 \cdots k_{N-1}} \pi_j \Pi_{\ell=1}^N P_{k_{\ell -1}k_\ell}^{(\ell)} \notag\\
  &= \sum_j \pi_j P_{ji}, \notag
\end{align}
where $\sum_{k_1 k_2 \cdots k_{N-1}} \Pi_{\ell=1}^N P_{k_{\ell -1}k_\ell}^{(\ell)} = P_{ji}$, and $j$ runs over all the states in the state space.
Therefore, using our algorithm for the local variables ensures the global balance of the total system.

It is noteworthy that the present algorithm reduces to the Suwa–Todo algorithm~\cite{SuwaT2010} when we set $s=\max_i \pi_i=:\pi_{\rm max}$, as in Fig.~\ref{fig:shift}(c).
The present extension to an arbitrary periodic shift using the tower picture makes it straightforward to understand that a rejection-free solution, that is, $v_{ii}=\pi_i P_{ii}=0$ $\forall i$, can be obtained for $\pi_{\rm max} \leq s \leq F_n - \pi_{\rm max}$ if $\pi_{\rm max} \leq F_n/2$, and there is no rejection-free solution otherwise.
In particular, the choice of $s=F_n/2$ always minimizes the overlap between the same color ranges in the original and shifted towers, providing an optimal transition kernel with the minimized rejection probability.
This is graphically clear in the periodic tower shift shown in Fig.~\ref{fig:shift}.
The transition kernel for $s=F_n/2$ is reversible because any point in the tower returns to its original position after two shifts of $s=F_n/2$.
In contrast, for $s \neq F_n/2$, the transition kernel is nonreversible due to the lack of this property and the resulting asymmetric transition between the states.
The present algorithm enables us to study the effect of the rejection probability by varying the shift parameter $s$.

In simulations, we calculate all the needed transition probabilities and prepare a lookup table of the probabilities before Monte Carlo sampling.
For nearest-neighbor interactions, the transition matrix of a local variable is determined by the state of the neighboring variables (environment).
The number of probabilities to store in memory is $n^{z+2}$, where $n$ is the number of local states, assumed the same for all local variables, and $z$ is the coordination number.
In the absence of translational symmetry, such as for interactions with randomness in spin-glass systems, the transition matrix depends on the site, and a prefactor of $N$ is multiplied, where $N$ is the number of sites.
Nevertheless, we can prepare a lookup table easily for systems with translational symmetry and also for systems without translational symmetry up to a fairly large system size.
In addition, if the transition matrix is sparse, the needed memory can be significantly reduced.
The shape of the environment for the local variable update depends on the local connectivity and the type of interaction the system has.
The coordination number may depend on the site in a random graph.
Not only two-body but many-body interactions can be treated on the same footing by expanding the environment.
The cost of preparing a lookup table is usually negligible, being much lower than the Monte Carlo sampling cost.
During simulations, we use Walker's alias method to efficiently generate a random event~\cite{FukuiT2009,HoritaST2017}, in which the computation time is $O(1)$ independent of the number of possible states $n$.
Therefore, when a lookup table can be prepared in memory, our algorithm is not extra costly compared to the standard Metropolis algorithm.
A reduction in autocorrelation time directly results in a reduction in the wall clock time.

\begin{figure}
  \begin{center}
\includegraphics[width=\columnwidth]{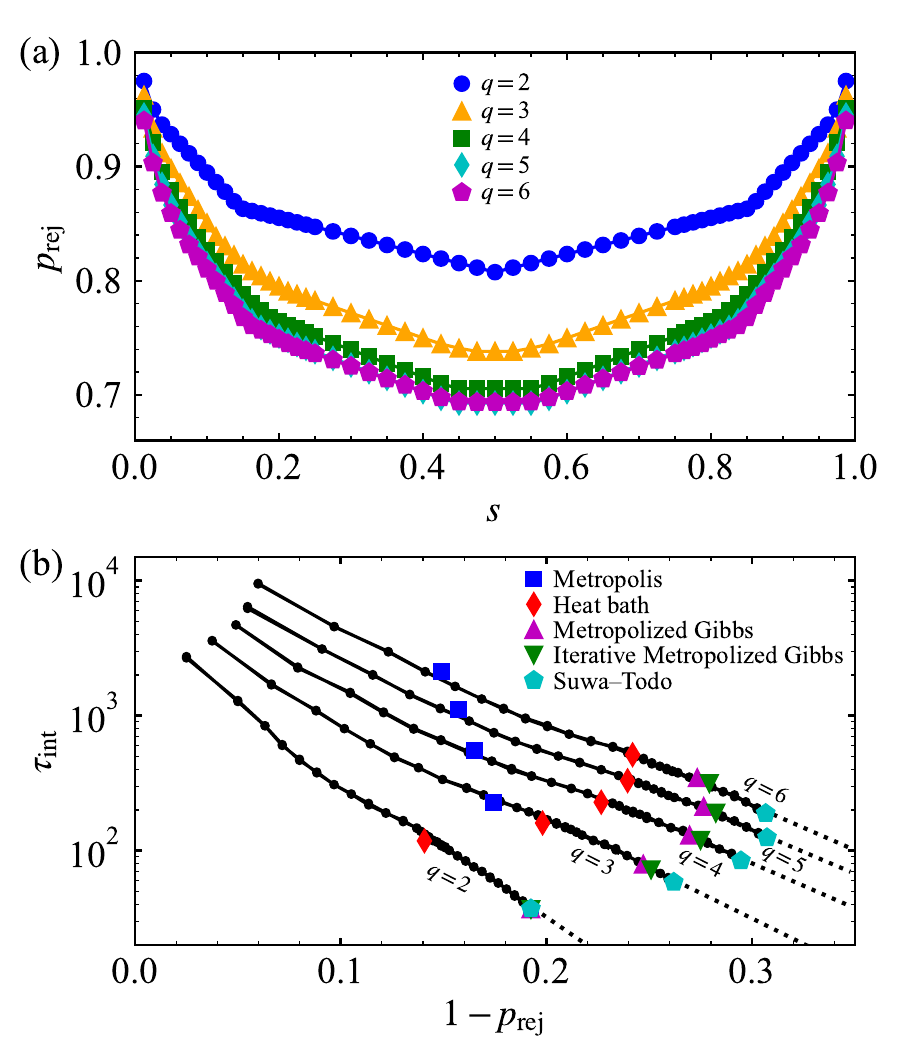}
\caption{\label{fig:2d}(a) Rejection rate for each shift parameter $s$ in the square-lattice $q$-state Potts model for $L=32$ and $q=2,3,4,5$ and 6 at the transition temperature. 
(b) Integrated autocorrelation time of the order parameter obtained using $s$ (circles) plotted in (a) and several algorithms: the Metropolis algorithm (squares), the heat bath algorithm (diamonds), the Metropolized Gibbs sampler (upper triangles), the iterative Metropolized Gibbs sampler (lower triangles), and the Suwa--Todo algorithm (pentagons).
The horizontal axis is $1- p_{\rm rej}$, where $p_{\rm rej}$ is the rejection rate measured during the simulations.
Each spin of the system is updated sequentially in a fixed order.
The autocorrelation times in the compared algorithms almost fall on a single curve obtained using the present algorithm with different $s$.
The dashed lines show the fitted exponential functions:
$\tau_{\rm int} = a e^{- b p_{\rm rej}}$, where $a$ and $b$ are the fitting parameters.
The decay factor $b$ is estimated to be $24$ for $q=2$, $17$ for $q=3$, and $15$ for $q=4,5,6$.
The error bars are smaller than the symbol sizes.
}
  \end{center}
\end{figure}

\section{Results}
\label{sec:result}
We examine how the rejection process affects the computational efficiency of the MCMC method using the $q$-state ferromagnetic Potts model~\cite{Wu1982}:
\begin{equation}
    H = - \sum_{\langle ij \rangle} \delta_{\sigma_i \sigma_j},
    \label{H}
\end{equation}
where $\langle ij \rangle$ runs over all the pairs of nearest neighbor sites, $\sigma_i = 1,2, \dots, q$ is the ``spin" at site $i$, and $\delta$ is the Kronecker delta.
This model shows the continuous phase transition for $q \leq 4$ and the first-order transition for $q > 4$ in the square lattice, while showing the continuous transition for $q = 2$ and the first-order transition for $q \geq 3$ in the simple cubic lattice.
The transition temperature is known in the square lattice model as $T_{\rm c}=1/\ln(1+\sqrt{q})$~\cite{Wu1982} and numerically estimated in the simple cubic lattice model to be $1/T_{\rm c}= 0.44330910(6)$~\cite{Deng2003}, $0.550565(10)$~\cite{Janke1997}, and $0.628616(16)$~\cite{Arnold1997} for $q=2,3$ and 4, respectively.
The numbers in parentheses indicate the statistical error in the preceding digit.
Note that the $q=2$ and $3$ models are equivalent to the Ising and the three-state clock ($\mathbb{Z}_3$) models, respectively, and the $q=4$ model is a special case of the Ashkin–Teller model~\cite{AshkinT1943,Baxter1982}.
Experimental realizations of the Potts model have been proposed and discussed~\cite{Domany1982}.

We calculate the order parameter defined by
\begin{equation}
    O^2 = \frac{q-1}{q} {\bm m}^2,
    \label{eq:op}
\end{equation}
where
\begin{align}
    {\bm m} &= (m_1, m_2, \dots, m_q) \\
    m_j &= \frac{1}{N (q-1)} \left\langle \sum_k ( q \delta_{j \sigma_k} - 1)\right\rangle,
\end{align}
$N$ is the number of sites, and the bracket $\langle \cdot \rangle$ denotes the Monte Carlo average.
To quantify the sampling efficiency of the Monte Carlo updates, we estimate the integrated autocorrelation time~\eqref{eq:tauint} of the order parameter~\eqref{eq:op}, using the asymptotic relation $\tau_{\rm int} = \sigma^2 / 2\bar{\sigma}^2$, where $\sigma^2$ and $\bar{\sigma}^2$ are the mean squared error, or the square of the statistical error, calculated from the binning analysis and without binning~\cite{Berg2004}.
For each Markov chain, we ran $2^{27}$ Monte Carlo steps, each consisting of $N$ local spin updates, and discarded the first half for the thermalization (burn-in) process.
We then calculated the integrated autocorrelation time and took the average over 128 independent Markov chains, enabling us to estimate the error bar reliably.

Varying the shift parameter introduced in Sec.~\ref{sec:one}, we compare the autocorrelation time in the present algorithm and several reference algorithms: the Metropolis algorithm~\cite{Metropolis1953}, the heat bath algorithm~\cite{Creutz1980,Geman1984}, the Metropolized Gibbs sampler~\cite{Liu1996}, the iterative Metropolized Gibbs sampler~\cite{Frigessi1992}, and the Suwa--Todo algorithm~\cite{SuwaT2010}.
For each local spin update, the next state is stochastically chosen from $n$ $(=q)$ states.
If the next state is the same as the current state, we consider it as a rejection process.
In the Metropolis algorithm, a state different from the current state was randomly proposed with probability $1/(n-1)$ and accepted/rejected by the standard Metropolis filter.
In using the present algorithm, we fix the order of local spin states, which are equivalent to each other due to the permutation symmetry of the Potts model.
Although our algorithm is described in Sec.~\ref{sec:one} without normalizing the weights, here we assume the normalization for simplicity: $F_n = \sum_{i=1}^n \pi_i = 1$, so the shift parameter $s$ can take $0 < s < 1$.
Note that the Metropolized Gibbs samplers and the Suwa--Todo algorithm reduce to the Metropolis algorithm for $q=2$.

\begin{figure}
  \begin{center}
\includegraphics[width=\columnwidth]{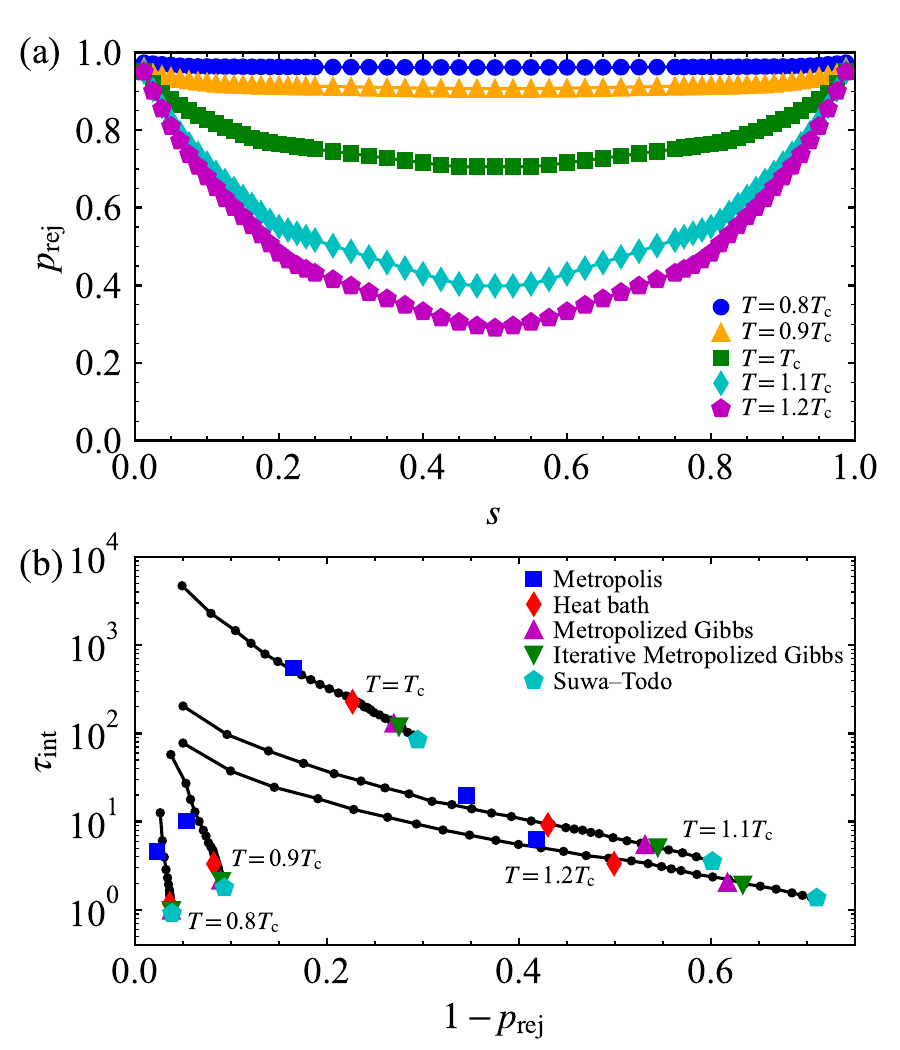}
\caption{\label{fig:T}(a) Rejection rate for each shift parameter $s$ in the square-lattice four-state Potts model for $L=32$ and at $T/T_{\rm c}=0.8,0.9,1.0,1.1$ and $1.2$, where $T_{\rm c}$ is the transition temperature. 
(b) Integrated autocorrelation time of the order parameter obtained using $s$ (circles) plotted in (a) and several reference algorithms, as in Fig.~\ref{fig:2d}.
The autocorrelation time almost exponentially decreases with reducing the rejection rate at all temperatures studied, and the autocorrelation times of the compared algorithms are in agreement with those of our algorithm with the corresponding rejection rates.
}
  \end{center}
\end{figure}

Figure~\ref{fig:2d}(a) shows the rejection rate measured during the simulations of the square-lattice $q$-state Potts model at the transition temperature for each shift parameter $s$ and $q=2,3,4,5$ and 6.
The system length was set to $L=32$ and $N=L^2$.
Here, we used the sequential spin update, in which the spins were swept sequentially in a fixed order.
The rejection rate is symmetric to $s=0.5$, as shown in Fig.~\ref{fig:2d}(a).
This symmetry is expected since the order of the local spin states is arbitrary in the Potts model, and the current choice is equivalent to the inverse order.
A similar symmetric behavior of the transition rate is also expected for other models, e.g., the clock model~\cite{Wu1982}.

Figure~\ref{fig:2d}(b) shows the integrated autocorrelation time of the order parameter calculated using different $s$ and reference algorithms.
Noticeably, the autocorrelation time decreases exponentially as the rejection rate is reduced.
We fitted the autocorrelation time to an exponential function and obtained large decay factors, as shown in the caption.
The decay factor is almost the same for $q=4,5$ and 6, implying a universal property for $q \geq 4$ in two dimensions.
Interestingly, the autocorrelation times of the reference algorithms almost fall on a single curve obtained using the present algorithm with different $s$.
This striking result indicates that the rejection probability almost determines the efficiency of Monte Carlo samplings.
The present algorithm with $s=0.5$ ensures the detailed balance, while the Suwa–Todo algorithm does not.
However, the two algorithms produce almost the same autocorrelation time, suggesting that the stochastic flow created locally in the state space is insignificant.
We are likely to need to use global state variables to induce effective stochastic flows, as the lifting technique considers~\cite{TuritsynCV2011,FernandesW2011,Schram2015,Vucelja2016,Faizi2020}.
We also calculated the integrated autocorrelation time of the total energy and found it to be shorter than that of the order parameter.
The ratio of the two autocorrelation times is almost constant, independent of $s$.
Thus, the present results are not specific to the order parameter but are expected to be common among many quantities.
Using a different system size only changes the prefactor of the autocorrelation time.
Our main results do not depend on the system size
(see~\ref{App:L} for the size dependence).

Figure~\ref{fig:T} shows the rejection rate and the integrated autocorrelation time of the order parameter for $q=4$ and at $T/T_{\rm c}=0.8,0.9,1.0,1.1$ and $1.2$.
Reducing the rejection rate leads to an exponential decrease in the autocorrelation time not only at the transition point but also at all temperatures studied.
Noticeably, the decay factor of the exponential function, which is $b$ in the relation $\tau_{\rm int} \sim a e^{- b p_{\rm rej}}$, increases with decreasing temperature.
The squared order parameter defined by Eq.~\eqref{eq:op} is $0.948435(2)$, $0.871663(6)$, $0.4627(3)$, $0.015438(4)$, and $0.007546(1)$ at $T/T_{\rm c}=0.8,0.9,1.0,1.1$ and $1.2$, respectively.
Thus, $T/T_{\rm c}=0.8$ is deep in the ordered phase and $T/T_{\rm c}=1.2$ is in the disordered phase, away from the transition point.
As shown in Fig.~\ref{fig:T}(b), the autocorrelation times of the reference algorithms agree with those of our algorithm with the corresponding rejection rates at all temperatures.
The agreement looks the best at the transition point.
Nevertheless, the universal relation between the rejection rate and the autocorrelation time is valid not only at the transition point but also in a wide temperature range.
Note that the ordered configuration becomes more stable at a lower temperature, and the rejection rate of the single spin update consequently becomes higher.

\begin{figure}
  \begin{center}
\includegraphics[width=\columnwidth]{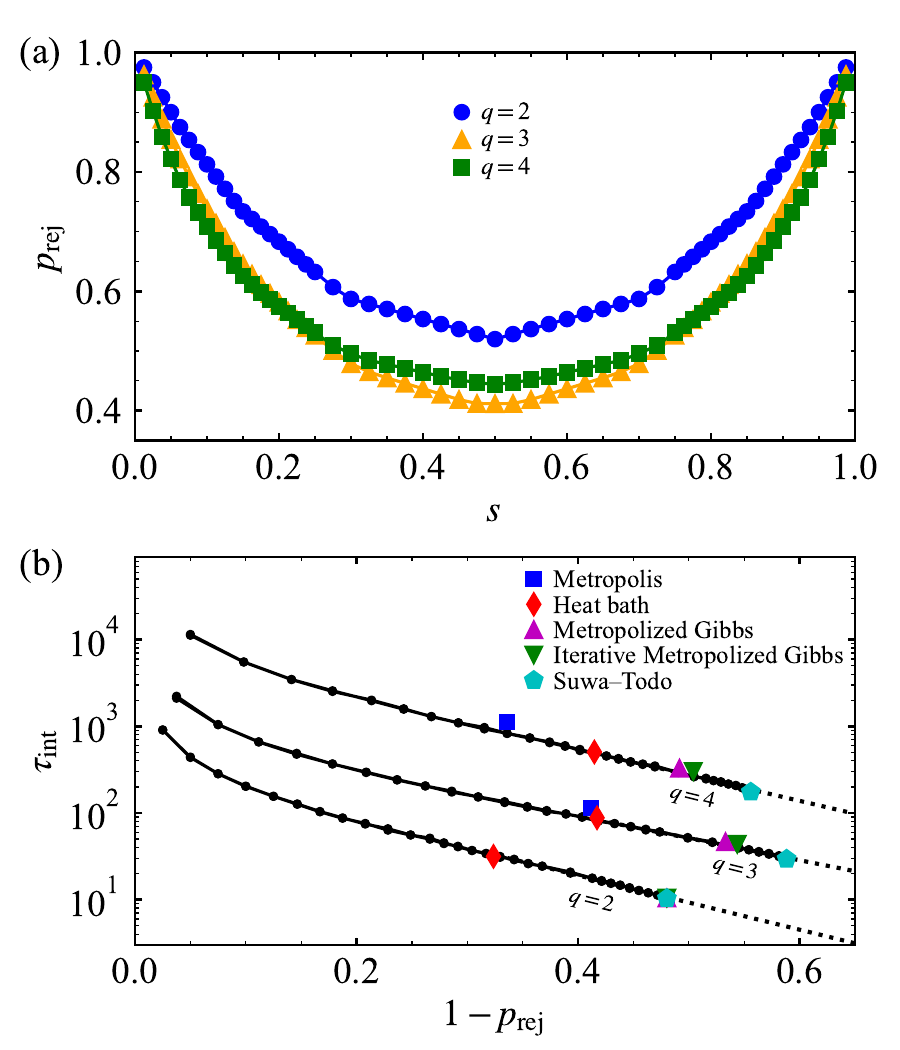}
\caption{\label{fig:3d}(a) Rejection rate for each shift parameter $s$ in the simple-cubic-lattice $q$-state Potts model for $L=12$ and $q=2,3$ and 4 at the transition temperature. 
(b) Integrated autocorrelation time of the order parameter obtained using $s$ (circles) plotted in (a) and several algorithms, as in Fig.~\ref{fig:2d}.
Although the Metropolis algorithm shows a slight deviation from the curve for $q=3$ and $4$, the autocorrelation times in the reference algorithms are almost consistent with the curve obtained using the present algorithm with different $s$.
The dashed lines show fitted exponential functions, and the decay factor is estimated to be $7.2$ for $q=2$, $5.9$ for $q=3$, and $6.9$ for $q=4$.}
  \end{center}
\end{figure}

To check whether the square lattice model is special, we calculate the rejection rate and the autocorrelation time in the simple cubic lattice model at the transition temperature for $q=2,3$ and 4, as shown in Fig.~\ref{fig:3d}.
The system length was set to $L=12$ and $N=L^3$.
As the rejection probability is reduced, the autocorrelation time decreases exponentially, similar to the case for the square lattice model.
Although the Metropolis algorithm shows a slight deviation, the autocorrelation times in the reference algorithms almost collapse onto a single curve obtained using the present algorithm with different shift parameters, as in the square lattice model.
Therefore, the present exponential improvement is expected to be universal among various systems.

\begin{figure}[b]
  \begin{center}
\includegraphics[width=\columnwidth]{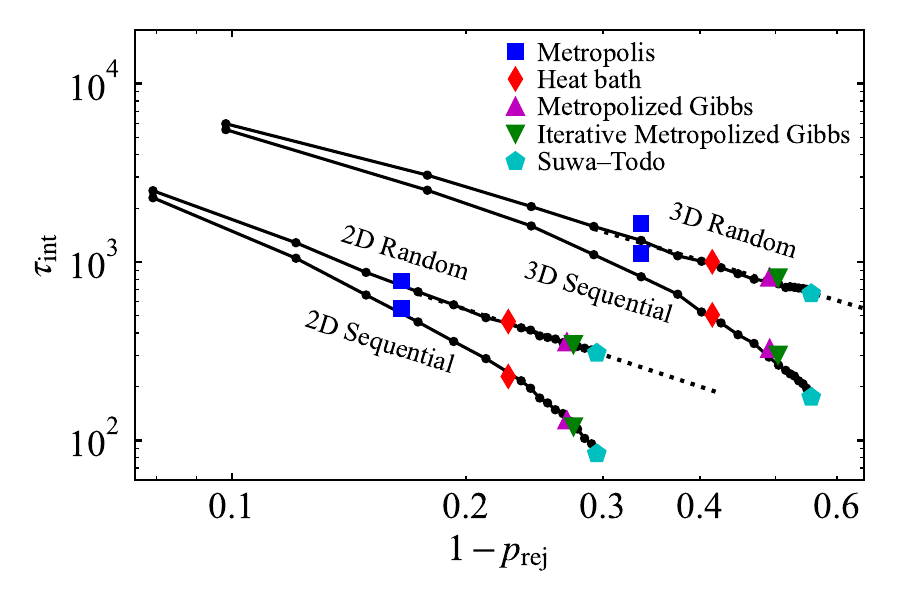}
\caption{\label{fig:comp}Integrated autocorrelation time of the order parameter obtained using the random and sequential spin updates in the square (2D) and cubic (3D) lattice four-state Potts model at the transition temperature.
The system size is set to $L=32$ and $12$ for the square and cubic lattice models, respectively.
The dashed lines show fitted power-law functions for the random spin update: $\tau_{\rm int}=\alpha (1-p_{\rm rej})^{-\gamma}$, where $\alpha$ and $\gamma$ are the fitting parameters.
The powers $\gamma$ are estimated to be $1.4$ and $1.3$ for the 2D and 3D cases, respectively.
}
  \end{center}
\end{figure}

Instead of the sequential spin update, we also test the random spin update, in which a spin (or a site) is randomly chosen for every update.
Figure~\ref{fig:comp} shows the autocorrelation time in the square and cubic lattice models with $q=4$ obtained using the random and sequential spin updates.
In contrast to the case of the sequential spin update, the autocorrelation time in the random spin update is well fitted to a power-law function: $\tau_{\rm int} \propto (1-p_{\rm rej})^{-\gamma}$.
The estimated powers $\gamma$ for the two- and three-dimensional models are noticeably close to each other, as shown in the caption of Fig.~\ref{fig:comp}.
Although the random spin update is more analytically tractable, the sequential spin update is superior to the random spin update as a sampling method.

\section{Conclusions}
\label{sec:con}
We have proposed a one-parameter rejection control transition kernel for the MCMC method and investigated the rejection effect on sampling efficiency using the $q$-state ferromagnetic Potts model in square and simple cubic lattices.
In the sequential spin update, the integrated autocorrelation time of the order parameter decreases exponentially as the rejection probability is reduced.
The decay factor for $q=4,5$ and 6 is almost the same in two dimensions, which implies a universal property of the system.
In contrast, the autocorrelation time decreases algebraically in the random spin update.
The significant difference between the rejection dependence of the sequential and random spin updates highlights the nontrivial relation between the matrix elements and the eigenvalues of the transition matrix in the many-body system.
Although the present and reference algorithms construct considerably different transition matrix structures, the autocorrelation times of the compared algorithms fall on almost a single curve as a function of the rejection rate.
Our finding indicates that reducing the rejection probability is essential for an efficient Monte Carlo update, and the rejection rate almost determines the sampling efficiency.
The present result establishes a straightforward guiding principle for the optimal transition kernel of discrete-variable interacting systems: Reducing the rejection probability is expected to produce efficient updates in various MCMC methods.
Furthermore, it is possible for the present algorithm to optimize other quantities, such as net stochastic flows, in combination with the lifting technique~\cite{Suwa2022}.

In the present paper, we have investigated the rejection effect on a local update.
It is interesting to study a guiding principle for nonlocal or cluster updates.
If the cluster shape is fixed, like a simple simultaneous update of fixed multiple spins, the resulting udpate is considered fundamentally local, and our guiding principle is expected to be applied.
If a cluster is constructed stochastically and can be extensive, the cluster flip results in a nonlocal update.
It is desirable to perform as nonlocal a cluster update as possible with as high a probability as possible.
The Swendsen-Wang~\cite{SwendsenW1987} and Wolff~\cite{Wolff1989} algorithms perform nonlocal cluster flips with probability one in several systems, but their extension to general models is highly nontrivial.
The search for a practical guiding principle for nonlocal updates in general cases is an intriguing future problem.
Recently, machine learning techniques based on autoregressive neural networks have been introduced to make a smart proposal distribution~\cite{Nicoli2020,McNaughton2020}.
A global update and almost uncorrelated sampling can be achieved using a trained generative neural network with a reasonably high acceptance probability.
Our approach can be combined with autoregressive algorithms to further reduce the rejection rate.
It is of great interest to apply the present algorithm to other systems and generalize the argument to nonlocal updates and continuous variable systems in the near future.

\section*{Acknowledgments}
The author is grateful to Synge Todo for the fruitful discussions.
Simulations were performed using computational resources from the Supercomputer Center at the Institute for Solid State Physics, the University of Tokyo.
This work was supported by the JSPS KAKENHI Grants-In-Aid for Scientific Research (No. 22K03508).

\appendix
\section{Derivation of the analytical expression}
\label{App:v}
We show here the derivation of the analytical expression of our method introduced in Sec.~\ref{sec:one}.
As explained above in the section, we set the stochastic flow to the overlap between the original and the shifted towers.
The allocation of stochastic flow is given by Eq.~\eqref{v_ij}.
The first term of the equation is the overlap between the ranges $[F_{j-1}, F_j]$ and $[F_{i-1} + s, F_i + s]$ with the shift parameter $s$.
Considering each possible case, the overlap is represented by
\begin{widetext}
\begin{subequations}
    \newcommand{\wedgesp}{\ \wedge \ } 
\begin{empheq}[left={v_{ij} = \empheqlbrace}]{align}
    0 & \qquad \text{if } F_{j-1} > F_i + s \\
    F_i - F_{j-1} + s & \qquad \text{if } F_j > F_i + s \wedgesp F_{j-1} < F_i + s \wedgesp F_{j-1} > F_{i-1}+s \\
    \pi_i & \qquad \text{if } F_j > F_i + s \wedgesp F_{j-1} < F_{i-1} + s \\
    \pi_j & \qquad \text{if } F_j < F_i + s \wedgesp F_{j-1} > F_{i-1} + s\\  
    F_j - F_{i-1} - s & \qquad \text{if } F_j < F_i + s \wedgesp F_j > F_{i-1} + s \wedgesp F_{j-1} < F_{i-1}+s\\    
    0 & \qquad \text{if } F_j < F_{i-1} + s .
\end{empheq}
\end{subequations}
Using $\Delta_{ij} = F_i - F_{j-1} + s$, Eq.~\eqref{delta}, we can rewrite these equations as
\begin{subequations}
    \newcommand{\wedgesp}{\ \wedge \ } 
\begin{empheq}[left={v_{ij} = \empheqlbrace}]{align}
    0 & \qquad \text{if } \Delta_{ij} < 0 \\
    \Delta_{ij} & \qquad \text{if } 0 < \Delta_{ij} < \pi_i \wedgesp \Delta_{ij} < \pi_j\\
    \pi_i & \qquad \text{if } \pi_i < \Delta_{ij} < \pi_j \\
    \pi_j & \qquad \text{if } \pi_j < \Delta_{ij} < \pi_i \\  
    \pi_i + \pi_j - \Delta_{ij} & \qquad \text{if } 0 < \pi_i + \pi_j - \Delta_{ij} < \pi_i \wedgesp \pi_i + \pi_j - \Delta_{ij} < \pi_j\\    
    0 & \qquad \text{if } \pi_i + \pi_j - \Delta_{ij} < 0
\end{empheq}
\end{subequations}
\end{widetext}
Combining these cases, we obtain
\begin{eqnarray}
  v_{ij} &=& \max(0, \ \min( \Delta_{ij}, \ \pi_i + \pi_j - \Delta_{ij}, \ \pi_i, \ \pi_j)). \notag\\
  \label{v_ij_1}
\end{eqnarray}
The second term of Eq.~\eqref{v_ij} is the overlap between the ranges $[F_{j-1}+F_n, F_j+F_n]$ and $[F_{i-1} + s, F_i + s]$.
The difference from the first case is only in adding $F_n$ to $F_j$ and $F_{j-1}$.
We can obtain the expression of the second case by replacing $\Delta_{ij}$ in Eq.~\eqref{v_ij_1} with $\Delta_{ij} - F_n$:
\begin{eqnarray}
  \!\!\!\! v_{ij} &=& \max(0, \ \min( \Delta_{ij} - F_n, \ \pi_i + \pi_j + F_n - \Delta_{ij}, \ \pi_i, \ \pi_j)). \notag\\
  \label{v_ij_2}
\end{eqnarray}
Adding the expressions in the two cases, Eqs.~\eqref{v_ij_1} and~\eqref{v_ij_2}, the analytical expression of the stochastic flow allocation is given by
\begin{eqnarray}
  \!\!\!\! v_{ij} &=& \max(0, \ \min( \Delta_{ij}, \ \pi_i + \pi_j - \Delta_{ij}, \ \pi_i, \ \pi_j)) \\
  && \hspace{-3.5mm} + \max(0, \ \min( \Delta_{ij} - F_n, \ \pi_i + \pi_j + F_n - \Delta_{ij}, \ \pi_i, \ \pi_j)), \notag
\end{eqnarray}
which is Eq.~\eqref{v_ij}.

\begin{figure}
  \begin{center}
\includegraphics[width=\columnwidth]{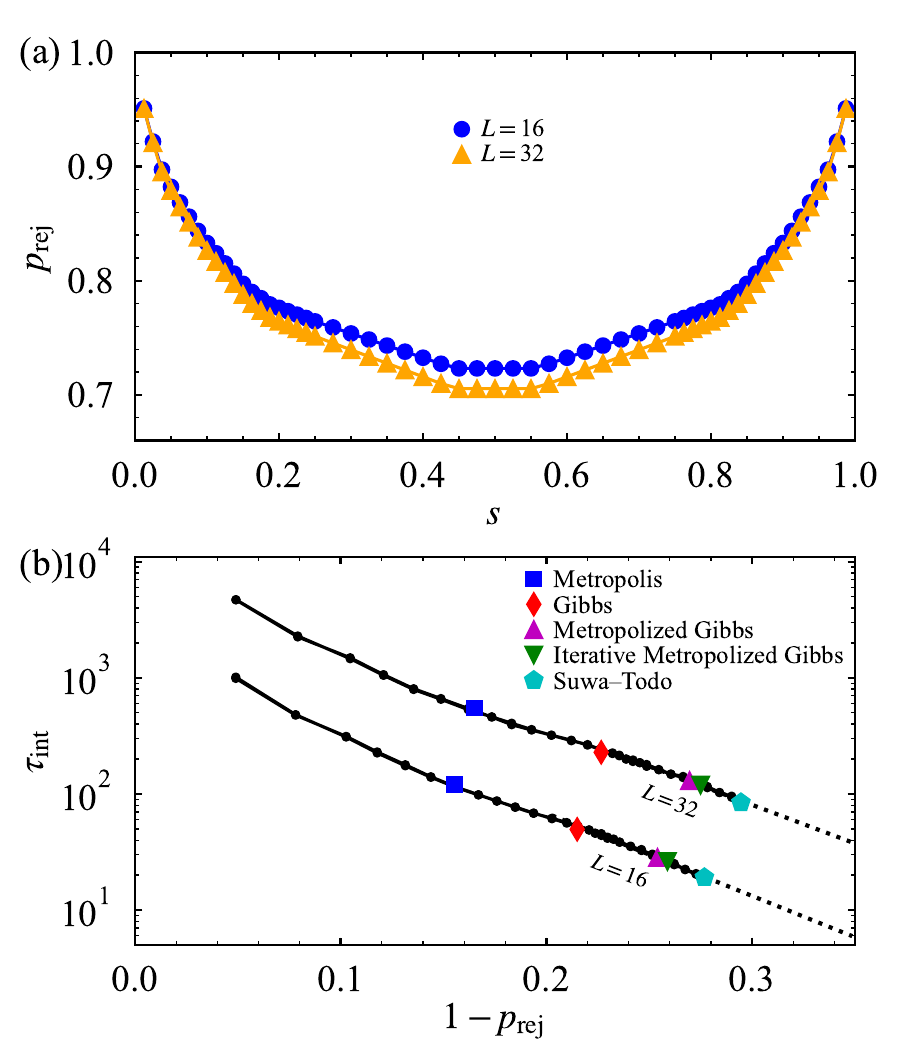}
\caption{\label{fig:L}(a) Rejection rate for each shift parameter $s$ and (b) integrated autocorrelation time of the order parameter in the square-lattice four-state Potts model for $L=16$ and $32$ at the transition temperature.
}
  \end{center}
\end{figure}

\section{System size dependence}
\label{App:L}
To check the size dependence of the main result, we calculate the rejection rate and the integrated autocorrelation time of the order parameter in the square-lattice four-state Potts model for $L=16$ and $32$.
As shown in Fig.~\ref{fig:L}(a), the rejection rate is slightly higher for $L=16$ than for $L=32$.
This is because ordered configurations appear more often in smaller system sizes, and the single spin flip from ordered configurations is more likely to be rejected.
The rejection rate for each $s$ should converge for large system sizes, and we expect the rate for $L=32$ to be close to the thermodynamic limit.
The autocorrelation time increases with $L$ at the transition point, given by $\tau_{\rm int} \propto L^{z}$ at a critical point, where $z$ here is the dynamic critical exponent, and $\tau_{\rm int} \propto e^{2\sigma L^{d-1}}$ at a first-order transition point, where $\sigma$ is the interface tension and $d$ is the dimension of the system.
However, the ratio of the autocorrelation times for $L=16$ and $32$ is almost independent of the update algorithm and the parameter $s$, as shown in Fig.~\ref{fig:L}(b).
In other words, the ratio of autocorrelation times for different algorithms is almost independent of the system size.
We expect that this should also be the case for other values of $q$.
Therefore, our main conclusions drawn from the results of $L=32$ should not depend on the system size.

\bibliography{main}

\begin{thebibliography}{39}%
\makeatletter
\providecommand \@ifxundefined [1]{%
 \@ifx{#1\undefined}
}%
\providecommand \@ifnum [1]{%
 \ifnum #1\expandafter \@firstoftwo
 \else \expandafter \@secondoftwo
 \fi
}%
\providecommand \@ifx [1]{%
 \ifx #1\expandafter \@firstoftwo
 \else \expandafter \@secondoftwo
 \fi
}%
\providecommand \natexlab [1]{#1}%
\providecommand \enquote  [1]{``#1''}%
\providecommand \bibnamefont  [1]{#1}%
\providecommand \bibfnamefont [1]{#1}%
\providecommand \citenamefont [1]{#1}%
\providecommand \href@noop [0]{\@secondoftwo}%
\providecommand \href [0]{\begingroup \@sanitize@url \@href}%
\providecommand \@href[1]{\@@startlink{#1}\@@href}%
\providecommand \@@href[1]{\endgroup#1\@@endlink}%
\providecommand \@sanitize@url [0]{\catcode `\\12\catcode `\$12\catcode
  `\&12\catcode `\#12\catcode `\^12\catcode `\_12\catcode `\%12\relax}%
\providecommand \@@startlink[1]{}%
\providecommand \@@endlink[0]{}%
\providecommand \url  [0]{\begingroup\@sanitize@url \@url }%
\providecommand \@url [1]{\endgroup\@href {#1}{\urlprefix }}%
\providecommand \urlprefix  [0]{URL }%
\providecommand \Eprint [0]{\href }%
\providecommand \doibase [0]{http://dx.doi.org/}%
\providecommand \selectlanguage [0]{\@gobble}%
\providecommand \bibinfo  [0]{\@secondoftwo}%
\providecommand \bibfield  [0]{\@secondoftwo}%
\providecommand \translation [1]{[#1]}%
\providecommand \BibitemOpen [0]{}%
\providecommand \bibitemStop [0]{}%
\providecommand \bibitemNoStop [0]{.\EOS\space}%
\providecommand \EOS [0]{\spacefactor3000\relax}%
\providecommand \BibitemShut  [1]{\csname bibitem#1\endcsname}%
\let\auto@bib@innerbib\@empty
\bibitem [{\citenamefont {Newman}\ and\ \citenamefont
  {Barkema}(1999)}]{Newman1999}%
  \BibitemOpen
  \bibfield  {author} {\bibinfo {author} {\bibfnamefont {M.E.J.}\ \bibnamefont
  {Newman}}\ and\ \bibinfo {author} {\bibfnamefont {G.T.}\ \bibnamefont
  {Barkema}},\ }\href@noop {} {\emph {\bibinfo {title} {{Monte} {Carlo} Methods
  in Statistical Physics}}}\ (\bibinfo  {publisher} {Oxford University Press},\
  \bibinfo {year} {1999})\BibitemShut {NoStop}%
\bibitem [{\citenamefont {Landau}\ and\ \citenamefont
  {Binder}(2005)}]{Landau2005}%
  \BibitemOpen
  \bibfield  {author} {\bibinfo {author} {\bibfnamefont {D.~P.}\ \bibnamefont
  {Landau}}\ and\ \bibinfo {author} {\bibfnamefont {K.}~\bibnamefont
  {Binder}},\ }\href@noop {} {\emph {\bibinfo {title} {A Guide to {Monte}
  {Carlo} Simulations in Statistical Physics}}},\ \bibinfo {edition} {2nd}\
  ed.\ (\bibinfo  {publisher} {Cambridge University Press},\ \bibinfo {address}
  {Cambridge},\ \bibinfo {year} {2005})\BibitemShut {NoStop}%
\bibitem [{\citenamefont {Robert}\ and\ \citenamefont
  {Casella}(2004)}]{Robert2004}%
  \BibitemOpen
  \bibfield  {author} {\bibinfo {author} {\bibfnamefont {Christian~P.}\
  \bibnamefont {Robert}}\ and\ \bibinfo {author} {\bibfnamefont {George}\
  \bibnamefont {Casella}},\ }\href@noop {} {\emph {\bibinfo {title} {{Monte
  Carlo} Statistical Methods}}},\ \bibinfo {edition} {2nd}\ ed.\ (\bibinfo
  {publisher} {Springer},\ \bibinfo {address} {New York},\ \bibinfo {year}
  {2004})\BibitemShut {NoStop}%
\bibitem [{\citenamefont {Metropolis}\ \emph {et~al.}(1953)\citenamefont
  {Metropolis}, \citenamefont {Rosenbluth}, \citenamefont {Rosenbluth},
  \citenamefont {Teller},\ and\ \citenamefont {Teller}}]{Metropolis1953}%
  \BibitemOpen
  \bibfield  {author} {\bibinfo {author} {\bibfnamefont {Nicholas}\
  \bibnamefont {Metropolis}}, \bibinfo {author} {\bibfnamefont {Arianna~W.}\
  \bibnamefont {Rosenbluth}}, \bibinfo {author} {\bibfnamefont {Marshall~N.}\
  \bibnamefont {Rosenbluth}}, \bibinfo {author} {\bibfnamefont {Augusta~H.}\
  \bibnamefont {Teller}}, \ and\ \bibinfo {author} {\bibfnamefont {Edward}\
  \bibnamefont {Teller}},\ }\bibfield  {title} {\enquote {\bibinfo {title}
  {Equation of state calculations by fast computing machines},}\ }\href
  {\doibase 10.1063/1.1699114} {\bibfield  {journal} {\bibinfo  {journal} {The
  Journal of Chemical Physics}\ }\textbf {\bibinfo {volume} {21}},\ \bibinfo
  {pages} {1087--1092} (\bibinfo {year} {1953})}\BibitemShut {NoStop}%
\bibitem [{\citenamefont {Creutz}(1980)}]{Creutz1980}%
  \BibitemOpen
  \bibfield  {author} {\bibinfo {author} {\bibfnamefont {Michael}\ \bibnamefont
  {Creutz}},\ }\bibfield  {title} {\enquote {\bibinfo {title} {Monte carlo
  study of quantized su(2) gauge theory},}\ }\href {\doibase
  10.1103/PhysRevD.21.2308} {\bibfield  {journal} {\bibinfo  {journal} {Phys.
  Rev. D}\ }\textbf {\bibinfo {volume} {21}},\ \bibinfo {pages} {2308--2315}
  (\bibinfo {year} {1980})}\BibitemShut {NoStop}%
\bibitem [{\citenamefont {Geman}\ and\ \citenamefont
  {Geman}(1984)}]{Geman1984}%
  \BibitemOpen
  \bibfield  {author} {\bibinfo {author} {\bibfnamefont {Stuart}\ \bibnamefont
  {Geman}}\ and\ \bibinfo {author} {\bibfnamefont {Donald}\ \bibnamefont
  {Geman}},\ }\bibfield  {title} {\enquote {\bibinfo {title} {Stochastic
  relaxation, gibbs distributions, and the bayesian restoration of images},}\
  }\href {\doibase 10.1109/TPAMI.1984.4767596} {\bibfield  {journal} {\bibinfo
  {journal} {IEEE Transactions on Pattern Analysis and Machine Intelligence}\
  }\textbf {\bibinfo {volume} {PAMI-6}},\ \bibinfo {pages} {721--741} (\bibinfo
  {year} {1984})}\BibitemShut {NoStop}%
\bibitem [{\citenamefont {Liu}(1996)}]{Liu1996}%
  \BibitemOpen
  \bibfield  {author} {\bibinfo {author} {\bibfnamefont {Jun~S.}\ \bibnamefont
  {Liu}},\ }\bibfield  {title} {\enquote {\bibinfo {title} {Metropolized
  independent sampling with comparisons to rejection sampling and importance
  sampling},}\ }\href {\doibase 10.1007/BF00162521} {\bibfield  {journal}
  {\bibinfo  {journal} {Statistics and Computing}\ }\textbf {\bibinfo {volume}
  {6}},\ \bibinfo {pages} {113--119} (\bibinfo {year} {1996})}\BibitemShut
  {NoStop}%
\bibitem [{\citenamefont {Frigessi}\ \emph {et~al.}(1992)\citenamefont
  {Frigessi}, \citenamefont {Hwang},\ and\ \citenamefont
  {Younes}}]{Frigessi1992}%
  \BibitemOpen
  \bibfield  {author} {\bibinfo {author} {\bibfnamefont {Arnoldo}\ \bibnamefont
  {Frigessi}}, \bibinfo {author} {\bibfnamefont {Chii-Ruey}\ \bibnamefont
  {Hwang}}, \ and\ \bibinfo {author} {\bibfnamefont {Laurent}\ \bibnamefont
  {Younes}},\ }\bibfield  {title} {\enquote {\bibinfo {title} {{Optimal
  Spectral Structure of Reversible Stochastic Matrices, Monte Carlo Methods and
  the Simulation of Markov Random Fields}},}\ }\href {\doibase
  10.1214/aoap/1177005652} {\bibfield  {journal} {\bibinfo  {journal} {The
  Annals of Applied Probability}\ }\textbf {\bibinfo {volume} {2}},\ \bibinfo
  {pages} {610 -- 628} (\bibinfo {year} {1992})}\BibitemShut {NoStop}%
\bibitem [{\citenamefont {Suwa}\ and\ \citenamefont {Todo}(2010)}]{SuwaT2010}%
  \BibitemOpen
  \bibfield  {author} {\bibinfo {author} {\bibfnamefont {Hidemaro}\
  \bibnamefont {Suwa}}\ and\ \bibinfo {author} {\bibfnamefont {Synge}\
  \bibnamefont {Todo}},\ }\bibfield  {title} {\enquote {\bibinfo {title}
  {Markov chain monte carlo method without detailed balance},}\ }\href
  {\doibase 10.1103/PhysRevLett.105.120603} {\bibfield  {journal} {\bibinfo
  {journal} {Phys. Rev. Lett.}\ }\textbf {\bibinfo {volume} {105}},\ \bibinfo
  {pages} {120603} (\bibinfo {year} {2010})}\BibitemShut {NoStop}%
\bibitem [{\citenamefont {Turitsyn}\ \emph {et~al.}(2011)\citenamefont
  {Turitsyn}, \citenamefont {Chertkov},\ and\ \citenamefont
  {Vucelja}}]{TuritsynCV2011}%
  \BibitemOpen
  \bibfield  {author} {\bibinfo {author} {\bibfnamefont {Konstantin~S.}\
  \bibnamefont {Turitsyn}}, \bibinfo {author} {\bibfnamefont {Michael}\
  \bibnamefont {Chertkov}}, \ and\ \bibinfo {author} {\bibfnamefont {Marija}\
  \bibnamefont {Vucelja}},\ }\bibfield  {title} {\enquote {\bibinfo {title}
  {Irreversible monte carlo algorithms for efficient sampling},}\ }\href
  {\doibase https://doi.org/10.1016/j.physd.2010.10.003} {\bibfield  {journal}
  {\bibinfo  {journal} {Physica D: Nonlinear Phenomena}\ }\textbf {\bibinfo
  {volume} {240}},\ \bibinfo {pages} {410--414} (\bibinfo {year}
  {2011})}\BibitemShut {NoStop}%
\bibitem [{\citenamefont {Fernandes}\ and\ \citenamefont
  {Weigel}(2011)}]{FernandesW2011}%
  \BibitemOpen
  \bibfield  {author} {\bibinfo {author} {\bibfnamefont {Heitor~C.M.}\
  \bibnamefont {Fernandes}}\ and\ \bibinfo {author} {\bibfnamefont {Martin}\
  \bibnamefont {Weigel}},\ }\bibfield  {title} {\enquote {\bibinfo {title}
  {Non-reversible monte carlo simulations of spin models},}\ }\href {\doibase
  https://doi.org/10.1016/j.cpc.2010.11.017} {\bibfield  {journal} {\bibinfo
  {journal} {Computer Physics Communications}\ }\textbf {\bibinfo {volume}
  {182}},\ \bibinfo {pages} {1856--1859} (\bibinfo {year} {2011})},\ \bibinfo
  {note} {computer Physics Communications Special Edition for Conference on
  Computational Physics Trondheim, Norway, June 23-26, 2010}\BibitemShut
  {NoStop}%
\bibitem [{\citenamefont {Schram}\ and\ \citenamefont
  {Barkema}(2015)}]{Schram2015}%
  \BibitemOpen
  \bibfield  {author} {\bibinfo {author} {\bibfnamefont {Raoul~D.}\
  \bibnamefont {Schram}}\ and\ \bibinfo {author} {\bibfnamefont {Gerard~T.}\
  \bibnamefont {Barkema}},\ }\bibfield  {title} {\enquote {\bibinfo {title}
  {Monte carlo methods beyond detailed balance},}\ }\href {\doibase
  https://doi.org/10.1016/j.physa.2014.06.015} {\bibfield  {journal} {\bibinfo
  {journal} {Physica A: Statistical Mechanics and its Applications}\ }\textbf
  {\bibinfo {volume} {418}},\ \bibinfo {pages} {88--93} (\bibinfo {year}
  {2015})},\ \bibinfo {note} {proceedings of the 13th International Summer
  School on Fundamental Problems in Statistical Physics}\BibitemShut {NoStop}%
\bibitem [{\citenamefont {Vucelja}(2016)}]{Vucelja2016}%
  \BibitemOpen
  \bibfield  {author} {\bibinfo {author} {\bibfnamefont {Marija}\ \bibnamefont
  {Vucelja}},\ }\bibfield  {title} {\enquote {\bibinfo {title} {Lifting—a
  nonreversible markov chain monte carlo algorithm},}\ }\href {\doibase
  10.1119/1.4961596} {\bibfield  {journal} {\bibinfo  {journal} {American
  Journal of Physics}\ }\textbf {\bibinfo {volume} {84}},\ \bibinfo {pages}
  {958--968} (\bibinfo {year} {2016})}\BibitemShut {NoStop}%
\bibitem [{\citenamefont {Faizi}\ \emph {et~al.}(2020)\citenamefont {Faizi},
  \citenamefont {Deligiannidis},\ and\ \citenamefont {Rosta}}]{Faizi2020}%
  \BibitemOpen
  \bibfield  {author} {\bibinfo {author} {\bibfnamefont {Fahim}\ \bibnamefont
  {Faizi}}, \bibinfo {author} {\bibfnamefont {George}\ \bibnamefont
  {Deligiannidis}}, \ and\ \bibinfo {author} {\bibfnamefont {Edina}\
  \bibnamefont {Rosta}},\ }\bibfield  {title} {\enquote {\bibinfo {title}
  {Efficient irreversible monte carlo samplers},}\ }\href {\doibase
  10.1021/acs.jctc.9b01135} {\bibfield  {journal} {\bibinfo  {journal} {Journal
  of Chemical Theory and Computation}\ }\textbf {\bibinfo {volume} {16}},\
  \bibinfo {pages} {2124--2138} (\bibinfo {year} {2020})}\BibitemShut {NoStop}%
\bibitem [{\citenamefont {Bernard}\ \emph {et~al.}(2009)\citenamefont
  {Bernard}, \citenamefont {Krauth},\ and\ \citenamefont
  {Wilson}}]{BernardKW2009}%
  \BibitemOpen
  \bibfield  {author} {\bibinfo {author} {\bibfnamefont {Etienne~P.}\
  \bibnamefont {Bernard}}, \bibinfo {author} {\bibfnamefont {Werner}\
  \bibnamefont {Krauth}}, \ and\ \bibinfo {author} {\bibfnamefont {David~B.}\
  \bibnamefont {Wilson}},\ }\bibfield  {title} {\enquote {\bibinfo {title}
  {Event-chain monte carlo algorithms for hard-sphere systems},}\ }\href
  {\doibase 10.1103/PhysRevE.80.056704} {\bibfield  {journal} {\bibinfo
  {journal} {Phys. Rev. E}\ }\textbf {\bibinfo {volume} {80}},\ \bibinfo
  {pages} {056704} (\bibinfo {year} {2009})}\BibitemShut {NoStop}%
\bibitem [{\citenamefont {Michel}\ \emph {et~al.}(2014)\citenamefont {Michel},
  \citenamefont {Kapfer},\ and\ \citenamefont {Krauth}}]{MichelKK2014}%
  \BibitemOpen
  \bibfield  {author} {\bibinfo {author} {\bibfnamefont {Manon}\ \bibnamefont
  {Michel}}, \bibinfo {author} {\bibfnamefont {Sebastian~C.}\ \bibnamefont
  {Kapfer}}, \ and\ \bibinfo {author} {\bibfnamefont {Werner}\ \bibnamefont
  {Krauth}},\ }\bibfield  {title} {\enquote {\bibinfo {title} {Generalized
  event-chain monte carlo: Constructing rejection-free global-balance
  algorithms from infinitesimal steps},}\ }\href {\doibase 10.1063/1.4863991}
  {\bibfield  {journal} {\bibinfo  {journal} {The Journal of Chemical Physics}\
  }\textbf {\bibinfo {volume} {140}},\ \bibinfo {pages} {054116} (\bibinfo
  {year} {2014})}\BibitemShut {NoStop}%
\bibitem [{\citenamefont {Peskun}(1973)}]{Peskun1973}%
  \BibitemOpen
  \bibfield  {author} {\bibinfo {author} {\bibfnamefont {P.~H.}\ \bibnamefont
  {Peskun}},\ }\bibfield  {title} {\enquote {\bibinfo {title} {{Optimum
  Monte-Carlo sampling using Markov chains}},}\ }\href {\doibase
  10.1093/biomet/60.3.607} {\bibfield  {journal} {\bibinfo  {journal}
  {Biometrika}\ }\textbf {\bibinfo {volume} {60}},\ \bibinfo {pages} {607--612}
  (\bibinfo {year} {1973})}\BibitemShut {NoStop}%
\bibitem [{\citenamefont {Suwa}\ and\ \citenamefont {Todo}(2012)}]{SuwaT2012}%
  \BibitemOpen
  \bibfield  {author} {\bibinfo {author} {\bibfnamefont {Hidemaro}\
  \bibnamefont {Suwa}}\ and\ \bibinfo {author} {\bibfnamefont {Synge}\
  \bibnamefont {Todo}},\ }\bibfield  {title} {\enquote {\bibinfo {title}
  {Geometric allocation approach for transition kernel of {Markov} chain},}\
  }in\ \href {\doibase 10.1515/9783110293586} {\emph {\bibinfo {booktitle}
  {Monte Carlo Methods and Applications: Proceedings of the 8th IMACS Seminar
  on Monte Carlo Methods}}}\ (\bibinfo {year} {2012})\ Chap.~\bibinfo {chapter}
  {23}, pp.\ \bibinfo {pages} {213--221}\BibitemShut {NoStop}%
\bibitem [{\citenamefont {Suwa}(2021)}]{Suwa2021}%
  \BibitemOpen
  \bibfield  {author} {\bibinfo {author} {\bibfnamefont {Hidemaro}\
  \bibnamefont {Suwa}},\ }\bibfield  {title} {\enquote {\bibinfo {title}
  {Geometric allocation approach to accelerating directed worm algorithm},}\
  }\href {\doibase 10.1103/PhysRevE.103.013308} {\bibfield  {journal} {\bibinfo
   {journal} {Phys. Rev. E}\ }\textbf {\bibinfo {volume} {103}},\ \bibinfo
  {pages} {013308} (\bibinfo {year} {2021})}\BibitemShut {NoStop}%
\bibitem [{\citenamefont {Chen}\ \emph {et~al.}(2012)\citenamefont {Chen},
  \citenamefont {Chen}, \citenamefont {Hwang},\ and\ \citenamefont
  {Pai}}]{Chen2012}%
  \BibitemOpen
  \bibfield  {author} {\bibinfo {author} {\bibfnamefont {Ting-Li}\ \bibnamefont
  {Chen}}, \bibinfo {author} {\bibfnamefont {Wei-Kuo}\ \bibnamefont {Chen}},
  \bibinfo {author} {\bibfnamefont {Chii-Ruey}\ \bibnamefont {Hwang}}, \ and\
  \bibinfo {author} {\bibfnamefont {Hui-Ming}\ \bibnamefont {Pai}},\ }\bibfield
   {title} {\enquote {\bibinfo {title} {On the optimal transition matrix for
  markov chain monte carlo sampling},}\ }\href {\doibase 10.1137/110832288}
  {\bibfield  {journal} {\bibinfo  {journal} {SIAM Journal on Control and
  Optimization}\ }\textbf {\bibinfo {volume} {50}},\ \bibinfo {pages}
  {2743--2762} (\bibinfo {year} {2012})}\BibitemShut {NoStop}%
\bibitem [{\citenamefont {Suwa}(2014)}]{Suwa2014}%
  \BibitemOpen
  \bibfield  {author} {\bibinfo {author} {\bibfnamefont {Hidemaro}\
  \bibnamefont {Suwa}},\ }\href@noop {} {\emph {\bibinfo {title} {Geometrically
  Constructed {Markov} Chain {Monte} {Carlo} Study of Quantum Spin-phonon
  Complex Systems}}},\ Springer Theses\ (\bibinfo  {publisher} {Springer},\
  \bibinfo {year} {2014})\BibitemShut {NoStop}%
\bibitem [{\citenamefont {Loison}\ \emph {et~al.}(2004)\citenamefont {Loison},
  \citenamefont {Qin}, \citenamefont {Schotte},\ and\ \citenamefont
  {Jin}}]{Loison2004}%
  \BibitemOpen
  \bibfield  {author} {\bibinfo {author} {\bibfnamefont {D.}~\bibnamefont
  {Loison}}, \bibinfo {author} {\bibfnamefont {C.~L.}\ \bibnamefont {Qin}},
  \bibinfo {author} {\bibfnamefont {K.~D.}\ \bibnamefont {Schotte}}, \ and\
  \bibinfo {author} {\bibfnamefont {X.~F.}\ \bibnamefont {Jin}},\ }\bibfield
  {title} {\enquote {\bibinfo {title} {Canonical local algorithms for spin
  systems: heat bath and hasting's methods},}\ }\href {\doibase
  10.1140/epjb/e2004-00332-5} {\bibfield  {journal} {\bibinfo  {journal} {The
  European Physical Journal B - Condensed Matter and Complex Systems}\ }\textbf
  {\bibinfo {volume} {41}},\ \bibinfo {pages} {395--412} (\bibinfo {year}
  {2004})}\BibitemShut {NoStop}%
\bibitem [{\citenamefont {Pollet}\ \emph {et~al.}(2004)\citenamefont {Pollet},
  \citenamefont {Rombouts}, \citenamefont {Van~Houcke},\ and\ \citenamefont
  {Heyde}}]{Pollet2004}%
  \BibitemOpen
  \bibfield  {author} {\bibinfo {author} {\bibfnamefont {Lode}\ \bibnamefont
  {Pollet}}, \bibinfo {author} {\bibfnamefont {Stefan M.~A.}\ \bibnamefont
  {Rombouts}}, \bibinfo {author} {\bibfnamefont {Kris}\ \bibnamefont
  {Van~Houcke}}, \ and\ \bibinfo {author} {\bibfnamefont {Kris}\ \bibnamefont
  {Heyde}},\ }\bibfield  {title} {\enquote {\bibinfo {title} {Optimal monte
  carlo updating},}\ }\href {\doibase 10.1103/PhysRevE.70.056705} {\bibfield
  {journal} {\bibinfo  {journal} {Phys. Rev. E}\ }\textbf {\bibinfo {volume}
  {70}},\ \bibinfo {pages} {056705} (\bibinfo {year} {2004})}\BibitemShut
  {NoStop}%
\bibitem [{\citenamefont {Todo}\ and\ \citenamefont {Suwa}(2013)}]{TodoS2013}%
  \BibitemOpen
  \bibfield  {author} {\bibinfo {author} {\bibfnamefont {S}~\bibnamefont
  {Todo}}\ and\ \bibinfo {author} {\bibfnamefont {H}~\bibnamefont {Suwa}},\
  }\bibfield  {title} {\enquote {\bibinfo {title} {Geometric allocation
  approaches in markov chain monte carlo},}\ }\href {\doibase
  10.1088/1742-6596/473/1/012013} {\bibfield  {journal} {\bibinfo  {journal}
  {Journal of Physics: Conference Series}\ }\textbf {\bibinfo {volume} {473}},\
  \bibinfo {pages} {012013} (\bibinfo {year} {2013})}\BibitemShut {NoStop}%
\bibitem [{\citenamefont {Fukui}\ and\ \citenamefont
  {Todo}(2009)}]{FukuiT2009}%
  \BibitemOpen
  \bibfield  {author} {\bibinfo {author} {\bibfnamefont {Kouki}\ \bibnamefont
  {Fukui}}\ and\ \bibinfo {author} {\bibfnamefont {Synge}\ \bibnamefont
  {Todo}},\ }\bibfield  {title} {\enquote {\bibinfo {title} {Order-n cluster
  monte carlo method for spin systems with long-range interactions},}\ }\href
  {\doibase https://doi.org/10.1016/j.jcp.2008.12.022} {\bibfield  {journal}
  {\bibinfo  {journal} {Journal of Computational Physics}\ }\textbf {\bibinfo
  {volume} {228}},\ \bibinfo {pages} {2629--2642} (\bibinfo {year}
  {2009})}\BibitemShut {NoStop}%
\bibitem [{\citenamefont {Horita}\ \emph {et~al.}(2017)\citenamefont {Horita},
  \citenamefont {Suwa},\ and\ \citenamefont {Todo}}]{HoritaST2017}%
  \BibitemOpen
  \bibfield  {author} {\bibinfo {author} {\bibfnamefont {Toshiki}\ \bibnamefont
  {Horita}}, \bibinfo {author} {\bibfnamefont {Hidemaro}\ \bibnamefont {Suwa}},
  \ and\ \bibinfo {author} {\bibfnamefont {Synge}\ \bibnamefont {Todo}},\
  }\bibfield  {title} {\enquote {\bibinfo {title} {Upper and lower critical
  decay exponents of ising ferromagnets with long-range interaction},}\ }\href
  {\doibase 10.1103/PhysRevE.95.012143} {\bibfield  {journal} {\bibinfo
  {journal} {Phys. Rev. E}\ }\textbf {\bibinfo {volume} {95}},\ \bibinfo
  {pages} {012143} (\bibinfo {year} {2017})}\BibitemShut {NoStop}%
\bibitem [{\citenamefont {Wu}(1982)}]{Wu1982}%
  \BibitemOpen
  \bibfield  {author} {\bibinfo {author} {\bibfnamefont {F.~Y.}\ \bibnamefont
  {Wu}},\ }\bibfield  {title} {\enquote {\bibinfo {title} {The potts model},}\
  }\href {\doibase 10.1103/RevModPhys.54.235} {\bibfield  {journal} {\bibinfo
  {journal} {Rev. Mod. Phys.}\ }\textbf {\bibinfo {volume} {54}},\ \bibinfo
  {pages} {235--268} (\bibinfo {year} {1982})}\BibitemShut {NoStop}%
\bibitem [{\citenamefont {Deng}\ and\ \citenamefont
  {Bl\"ote}(2003)}]{Deng2003}%
  \BibitemOpen
  \bibfield  {author} {\bibinfo {author} {\bibfnamefont {Youjin}\ \bibnamefont
  {Deng}}\ and\ \bibinfo {author} {\bibfnamefont {Henk W.~J.}\ \bibnamefont
  {Bl\"ote}},\ }\bibfield  {title} {\enquote {\bibinfo {title} {Simultaneous
  analysis of several models in the three-dimensional ising universality
  class},}\ }\href {\doibase 10.1103/PhysRevE.68.036125} {\bibfield  {journal}
  {\bibinfo  {journal} {Phys. Rev. E}\ }\textbf {\bibinfo {volume} {68}},\
  \bibinfo {pages} {036125} (\bibinfo {year} {2003})}\BibitemShut {NoStop}%
\bibitem [{\citenamefont {Janke}\ and\ \citenamefont
  {Villanova}(1997)}]{Janke1997}%
  \BibitemOpen
  \bibfield  {author} {\bibinfo {author} {\bibfnamefont {Wolfhard}\
  \bibnamefont {Janke}}\ and\ \bibinfo {author} {\bibfnamefont {Ramon}\
  \bibnamefont {Villanova}},\ }\bibfield  {title} {\enquote {\bibinfo {title}
  {Three-dimensional 3-state potts model revisited with new techniques},}\
  }\href {\doibase https://doi.org/10.1016/S0550-3213(96)00710-9} {\bibfield
  {journal} {\bibinfo  {journal} {Nuclear Physics B}\ }\textbf {\bibinfo
  {volume} {489}},\ \bibinfo {pages} {679--696} (\bibinfo {year}
  {1997})}\BibitemShut {NoStop}%
\bibitem [{\citenamefont {Arnold}\ and\ \citenamefont
  {Zhang}(1997)}]{Arnold1997}%
  \BibitemOpen
  \bibfield  {author} {\bibinfo {author} {\bibfnamefont {Peter}\ \bibnamefont
  {Arnold}}\ and\ \bibinfo {author} {\bibfnamefont {Yan}\ \bibnamefont
  {Zhang}},\ }\bibfield  {title} {\enquote {\bibinfo {title} {Monte carlo study
  of very weakly first-order transitions in the three-dimensional ashkin-teller
  model},}\ }\href {\doibase https://doi.org/10.1016/S0550-3213(97)00405-7}
  {\bibfield  {journal} {\bibinfo  {journal} {Nuclear Physics B}\ }\textbf
  {\bibinfo {volume} {501}},\ \bibinfo {pages} {803--837} (\bibinfo {year}
  {1997})}\BibitemShut {NoStop}%
\bibitem [{\citenamefont {Ashkin}\ and\ \citenamefont
  {Teller}(1943)}]{AshkinT1943}%
  \BibitemOpen
  \bibfield  {author} {\bibinfo {author} {\bibfnamefont {J.}~\bibnamefont
  {Ashkin}}\ and\ \bibinfo {author} {\bibfnamefont {E.}~\bibnamefont
  {Teller}},\ }\bibfield  {title} {\enquote {\bibinfo {title} {Statistics of
  two-dimensional lattices with four components},}\ }\href {\doibase
  10.1103/PhysRev.64.178} {\bibfield  {journal} {\bibinfo  {journal} {Phys.
  Rev.}\ }\textbf {\bibinfo {volume} {64}},\ \bibinfo {pages} {178--184}
  (\bibinfo {year} {1943})}\BibitemShut {NoStop}%
\bibitem [{\citenamefont {Baxter}(1982)}]{Baxter1982}%
  \BibitemOpen
  \bibfield  {author} {\bibinfo {author} {\bibfnamefont {Rodney~J.}\
  \bibnamefont {Baxter}},\ }\href@noop {} {\emph {\bibinfo {title} {Exactly
  Solved Models in Statistical Mechanics}}}\ (\bibinfo  {publisher} {ACADEMIC
  PRESS},\ \bibinfo {address} {San Diego},\ \bibinfo {year} {1982})\BibitemShut
  {NoStop}%
\bibitem [{\citenamefont {Domany}\ \emph {et~al.}(1982)\citenamefont {Domany},
  \citenamefont {Shnidman},\ and\ \citenamefont {Mukamel}}]{Domany1982}%
  \BibitemOpen
  \bibfield  {author} {\bibinfo {author} {\bibfnamefont {E}~\bibnamefont
  {Domany}}, \bibinfo {author} {\bibfnamefont {Y}~\bibnamefont {Shnidman}}, \
  and\ \bibinfo {author} {\bibfnamefont {D}~\bibnamefont {Mukamel}},\
  }\bibfield  {title} {\enquote {\bibinfo {title} {Type i {FCC}
  antiferromagnets in a magnetic field: a realisation of the q=3- and q=4-state
  potts models},}\ }\href {\doibase 10.1088/0022-3719/15/14/010} {\bibfield
  {journal} {\bibinfo  {journal} {Journal of Physics C: Solid State Physics}\
  }\textbf {\bibinfo {volume} {15}},\ \bibinfo {pages} {L495--L500} (\bibinfo
  {year} {1982})}\BibitemShut {NoStop}%
\bibitem [{\citenamefont {Berg}(2004)}]{Berg2004}%
  \BibitemOpen
  \bibfield  {author} {\bibinfo {author} {\bibfnamefont {Bernd~A.}\
  \bibnamefont {Berg}},\ }\href@noop {} {\emph {\bibinfo {title} {{Markov}
  Chain {Monte} {Carlo} Simulations and Their Statistical Analysis: With
  Web-based Fortran Code}}}\ (\bibinfo  {publisher} {World Scientific
  Publishing},\ \bibinfo {year} {2004})\BibitemShut {NoStop}%
\bibitem [{\citenamefont {Suwa}(2022)}]{Suwa2022}%
  \BibitemOpen
  \bibfield  {author} {\bibinfo {author} {\bibfnamefont {Hidemaro}\
  \bibnamefont {Suwa}},\ }\bibfield  {title} {\enquote {\bibinfo {title}
  {Lifted directed-worm algorithm},}\ }\href {\doibase
  10.1103/PhysRevE.106.055306} {\bibfield  {journal} {\bibinfo  {journal}
  {Phys. Rev. E}\ }\textbf {\bibinfo {volume} {106}},\ \bibinfo {pages}
  {055306} (\bibinfo {year} {2022})}\BibitemShut {NoStop}%
\bibitem [{\citenamefont {Swendsen}\ and\ \citenamefont
  {Wang}(1987)}]{SwendsenW1987}%
  \BibitemOpen
  \bibfield  {author} {\bibinfo {author} {\bibfnamefont {Robert~H.}\
  \bibnamefont {Swendsen}}\ and\ \bibinfo {author} {\bibfnamefont {Jian-Sheng}\
  \bibnamefont {Wang}},\ }\bibfield  {title} {\enquote {\bibinfo {title}
  {Nonuniversal critical dynamics in monte carlo simulations},}\ }\href
  {\doibase 10.1103/PhysRevLett.58.86} {\bibfield  {journal} {\bibinfo
  {journal} {Phys. Rev. Lett.}\ }\textbf {\bibinfo {volume} {58}},\ \bibinfo
  {pages} {86--88} (\bibinfo {year} {1987})}\BibitemShut {NoStop}%
\bibitem [{\citenamefont {Wolff}(1989)}]{Wolff1989}%
  \BibitemOpen
  \bibfield  {author} {\bibinfo {author} {\bibfnamefont {U.}~\bibnamefont
  {Wolff}},\ }\bibfield  {title} {\enquote {\bibinfo {title} {Collective
  {Monte} {Carlo} updating for spin systems},}\ }\href@noop {} {\bibfield
  {journal} {\bibinfo  {journal} {Phys. Rev. Lett.}\ }\textbf {\bibinfo
  {volume} {62}},\ \bibinfo {pages} {361} (\bibinfo {year} {1989})}\BibitemShut
  {NoStop}%
\bibitem [{\citenamefont {Nicoli}\ \emph {et~al.}(2020)\citenamefont {Nicoli},
  \citenamefont {Nakajima}, \citenamefont {Strodthoff}, \citenamefont {Samek},
  \citenamefont {M\"uller},\ and\ \citenamefont {Kessel}}]{Nicoli2020}%
  \BibitemOpen
  \bibfield  {author} {\bibinfo {author} {\bibfnamefont {Kim~A.}\ \bibnamefont
  {Nicoli}}, \bibinfo {author} {\bibfnamefont {Shinichi}\ \bibnamefont
  {Nakajima}}, \bibinfo {author} {\bibfnamefont {Nils}\ \bibnamefont
  {Strodthoff}}, \bibinfo {author} {\bibfnamefont {Wojciech}\ \bibnamefont
  {Samek}}, \bibinfo {author} {\bibfnamefont {Klaus-Robert}\ \bibnamefont
  {M\"uller}}, \ and\ \bibinfo {author} {\bibfnamefont {Pan}\ \bibnamefont
  {Kessel}},\ }\bibfield  {title} {\enquote {\bibinfo {title} {Asymptotically
  unbiased estimation of physical observables with neural samplers},}\ }\href
  {\doibase 10.1103/PhysRevE.101.023304} {\bibfield  {journal} {\bibinfo
  {journal} {Phys. Rev. E}\ }\textbf {\bibinfo {volume} {101}},\ \bibinfo
  {pages} {023304} (\bibinfo {year} {2020})}\BibitemShut {NoStop}%
\bibitem [{\citenamefont {McNaughton}\ \emph {et~al.}(2020)\citenamefont
  {McNaughton}, \citenamefont {Milo\ifmmode \check{s}\else
  \v{s}\fi{}evi\ifmmode~\acute{c}\else \'{c}\fi{}}, \citenamefont {Perali},\
  and\ \citenamefont {Pilati}}]{McNaughton2020}%
  \BibitemOpen
  \bibfield  {author} {\bibinfo {author} {\bibfnamefont {B.}~\bibnamefont
  {McNaughton}}, \bibinfo {author} {\bibfnamefont {M.~V.}\ \bibnamefont
  {Milo\ifmmode \check{s}\else \v{s}\fi{}evi\ifmmode~\acute{c}\else
  \'{c}\fi{}}}, \bibinfo {author} {\bibfnamefont {A.}~\bibnamefont {Perali}}, \
  and\ \bibinfo {author} {\bibfnamefont {S.}~\bibnamefont {Pilati}},\
  }\bibfield  {title} {\enquote {\bibinfo {title} {Boosting monte carlo
  simulations of spin glasses using autoregressive neural networks},}\ }\href
  {\doibase 10.1103/PhysRevE.101.053312} {\bibfield  {journal} {\bibinfo
  {journal} {Phys. Rev. E}\ }\textbf {\bibinfo {volume} {101}},\ \bibinfo
  {pages} {053312} (\bibinfo {year} {2020})}\BibitemShut {NoStop}%
\end{thebibliography}%

\end{document}